\documentclass[aps,pra,twocolumn,superscriptaddress]{revtex4-1}
\usepackage[utf8]{inputenc}
\usepackage{graphicx}
\usepackage{amsmath}
\usepackage{amsfonts}
\usepackage{url}

\begin{document}

\title{Long Distance Continuous-Variable Quantum Key Distribution with a Gaussian Modulation}

\author{Paul Jouguet}
\affiliation{Institut Telecom / Telecom ParisTech, CNRS LTCI, 46, rue Barrault, 75634 Paris Cedex 13, France}
\affiliation{SeQureNet, 23 avenue d'Italie, 75013 Paris, France}

\author{Sébastien Kunz-Jacques}
\affiliation{SeQureNet, 23 avenue d'Italie, 75013 Paris, France}

\author{Anthony Leverrier}
\affiliation{
%ICFO-Institut de Ci\`encies Fot\`oniques, 08860 Castelldefels (Barcelona), Spain
ICFO-Institut de Ciencies Fotoniques, 08860 Castelldefels (Barcelona), Spain
}

\setlength{\parskip}{0.5ex plus 0.25ex minus 0.1ex}

\newcommand{\vu}{\ensuremath{\mathbf{u}}}
\newcommand{\vv}{\ensuremath{\mathbf{v}}}
\newcommand{\vw}{\ensuremath{\mathbf{w}}}
\newcommand{\vx}{\ensuremath{\mathbf{x}}}
\newcommand{\vy}{\ensuremath{\mathbf{y}}}
\newcommand{\vz}{\ensuremath{\mathbf{z}}}
\newcommand{\myvector}[1]{\ensuremath{\mathbf{#1}}}
\date{\today}

\begin{abstract}
We designed high-efficiency error correcting codes allowing to extract an errorless secret key in a Continuous-Variable Quantum Key Distribution (CVQKD) protocol using a Gaussian modulation of coherent states and a homodyne detection. These codes are available for a wide range of signal-to-noise ratios on an Additive White Gaussian Noise Channel (AWGNC) with a binary modulation and can be combined with a multidimensional reconciliation method proven secure against arbitrary collective attacks. This improved reconciliation procedure considerably extends the secure range of CVQKD with a Gaussian modulation, giving a secret key rate of about $10^{-3}$ bit per pulse at a distance of 120 km for reasonable physical parameters.
\end{abstract}

\maketitle

\section{Introduction}
Quantum Key Distribution (QKD) \cite{SCA09} is the first real-life application of quantum information. It allows two distant parties, Alice and Bob, to establish an \textit{unconditional} \footnote{that is one does not have to make any assumption on the capacities of the eavesdropper (calculation power, knowledge of efficient algorithms, amount of memory\dots) to prove the security of the established key} secret key through the exchange of quantum states even in the presence of an eavesdropper, with the help of a classical auxiliary authenticated communication channel \footnote{For a discussion about how such a channel can be established in practice, refer to \cite{KUN11}}. The first QKD vendors, ID Quantique \cite{IDQ} and MagiQ Technologies \cite{MAG}, developed systems based on encoding the information on discrete variables such as the phase or the polarization of single photons. The limitation of these technologies is mainly due to the speed and the efficiency of the single photon detectors. Recently created companies, Quintessence Labs \cite{QUI} and SeQureNet \cite{SQN}, are developing a new generation of systems that allow to get rid of these limitations by encoding the information on continuous variables (CV) such as the quadratures of coherent states. 

In the standard protocol \cite{GRO02}, one needs to prepare Gaussian modulated coherent states and to measure them with a homodyne or a heterodyne \cite{WEE04} detection which requires only standard telecommunication parts. With the current proof techniques, using a Gaussian modulation is optimal as regards the theoretical secret key rate. In particular, security against collective attacks is well understood \cite{GAR06,NAV06}, even in the finite-size regime \cite{LEV10b}, and collective attacks are known to be asymptotically optimal \cite{REN09}. However, since the efficiency of the current reconciliation protocols for Gaussian variables drops dramatically in the regime of low signal-to-noise ratios (SNRs), new protocols using specific non-Gaussian modulations, either discrete \cite{LEV09} or continuous \cite{LEV11}, have been developed. The idea of these modulations is that they are compatible with high-performance error correction, making possible for the protocol parties to extract efficiently the information available in their raw data. This is in strong contrast with the Gaussian modulation for which no efficient reconciliation procedure was available until now. In theory, protocols with a non-Gaussian modulation therefore increase the achievable secure distance of CVQKD. They have, however, not yet been demonstrated experimentally. Indeed, for long distances, that is low transmission of the quantum channel, the optimal modulation variance is typically lower for non-Gaussian modulations (in particular for the four-state protocol \cite{LEV09}) than for a Gaussian modulation. This makes the design a stable continuous-variable system able to operate at large distances difficult. Even if this effect is mitigated for the eight-dimension protocol \cite{LEV11}, the modulation allowing for the largest variance remains the Gaussian one \cite{GRO02}. 

In this paper, we exhibit high-efficiency error correcting codes which can be combined with a multidimensional reconciliation scheme \cite{LEV08}. This allows, for the first time, to distill a secret key from a CVQKD protocol with a Gaussian modulation in the regime of very low SNR, and paves the way for future experimental demonstrations of CVQKD over much larger distances than the current record of 30 km \cite{LOD07, FOS09}.

In Section \ref{sec:gaussianrec}, we explain how the problem of reconciling Gaussian variables can be translated into a channel coding problem on the Binary Input Additive White Gaussian Noise Channel (BIAWGNC), for which we describe very low rate error correcting codes in Section \ref{sec:ldpcopt}. Combining these tools, we are able to efficiently reconcile data at low SNRs. Finally, we show in Section \ref{sec:practical} the consequences of these new developments on the performance of the Gaussian protocol over long distances.

\section{Theory of Reconciliation of Gaussian Variables}
\label{sec:gaussianrec}
The data reconciliation step is critical in CVQKD: the distance of the chosen error-correction scheme to the Shannon bound affects both the key rate and the range of the protocol. Of considerable importance is the problem of the reconciliation of correlated Gaussian variables. This is indeed the scenario considered in the GG02 protocol \cite{GRO02} where Alice's coherent states are modulated with a bivariate Gaussian distribution in phase space. Different approaches have been explored to increase the reconciliation efficiency for a Gaussian modulation, especially in the regime of low SNR.

A first approach called \textit{Slice Reconciliation} was proposed in \cite{VAN04,BLO05} and implemented in \cite{LOD07,FOS09} but the efficiency of this method currently limits the protocol range to about 30 km. Another method is to encode the information on the sign of the Gaussian modulated value. However, since we deal with centered Gaussian variables, the uncertainty on the sign increases at low SNRs because most values have small amplitude. Another class of protocols use \textit{post-selection} \cite{SIL02,NAM04,LAN05,HEI07} by working only with high-amplitude data but the security is not proven against general collective attacks.

In \cite{LEV08}, the idea of reducing the Gaussian variables reconciliation problem to the channel coding problem is introduced. One first uses a $d$-dimensional rotation to build a virtual channel close to the BIAWGNC from the physical Gaussian channel. This means that $d$ consecutive instances of the physical channel are mapped to $d$ approximate copies of a virtual BIAWGN channel, which are used to perform the error correction and eventually distill the actual secret. The final reconciliation efficiency one obtains with such a scheme depends on two things: 
\begin{itemize}
\item
The intrinsic efficiency of the error correcting code used on the virtual channel \textit{on the BIAWGNC}  (such an efficiency is given for example in Table \ref{tab:biawgnceff}).
\item
The quality of the approximation between the virtual channel and the BIAWGNC (for the scheme given in \cite{LEV08}, the quality of this approximation increases with the dimension $d$). 
\end{itemize}
One can therefore improve the reconciliation efficiency of the global scheme by working on two things: designing codes with higher efficiencies on the BIAWGNC and increasing the dimension of the scheme.

Let us now explain in more details the setting defined in \cite{LEV08}: Alice, the sender, and Bob, the receiver, are given two $n$-dimensional real vectors $\vx$ and $\vy$ and can use a public authenticated channel to agree on a common bit string $\vu$. For this, one of the parties (say Alice in the direct reconciliation scheme) sends to the other additional information describing a function $f$ such that $f(\vx) = \vu$; the other party (Bob) applies this function to his data to get $\vv := f(\vy)$; this way, a virtual communication channel with input $\vu$ and output $\vv$ is defined. The explicit construction of \cite{LEV08} aims at creating a virtual channel that is close to the BIAWGNC since very efficient codes are available for that channel.

Alice and Bob are given two $d$-uplets $\vx$ and $\vy$ corresponding to correlated Gaussian vectors (this is valid for CVQKD with a Gaussian modulation and a Gaussian optimal attack). This means that one can introduce constants $t, t'$ and $\sigma, \sigma'$ such that one has $\vy=t \vx+\vz$ with $\vx \sim \mathcal{N}(0,1)^d$, $\vz \sim \mathcal{N}(0,\sigma^2)^d$ in the direct reconciliation case and $\vx = t' \vy + \vz'$ with $\vy \sim \mathcal{N}(0,1+\sigma^2)^d$, $\vz'^d \sim \mathcal{N}(0,\sigma'^2)^d$ in the reverse reconciliation case. Since the two scenarios are similar, we consider without loss of generality only the direct reconciliation one here. Furthermore, up to a simple renormalization, one can fix $t=1$.

Alice chooses a random element $\vu \in \{-1/\sqrt{d},1/\sqrt{d}\}^d$ with the uniform distribution on the $d$-dimensional hypercube and sends $\myvector{r}=\vu . \vx^{-1}$ to Bob through the public channel (a ``multiplication'' and its inverse ``division'' operator are assumed to exist on $d$-dimensional vectors - more on this below). Then Bob computes $\vv:=\myvector{r} . \vy$. Let us analyse the noise $\vw$ on this virtual channel:
\begin{align*}
\vw & := \vv - \vu\\
& = \myvector{r} \vy - \vu\\
& = \vu . \vx^{-1}\left(\vx + \vz \right) - \vu\\
& = \vu \frac{\vz}{\vx} \sim \vu \frac{\vz}{||\vx||}
\end{align*}
where the last equality holds in law and is due to the spherical symmetry of the distributions of $\vz$ and $\vx$ and their independence. Since the norm of $\vx$ is transmitted, the channel considered is a Fading Channel with Known Side Information as defined in \cite{RIC08}, the fading coefficient being the norm of $\vx$, which follows a $\chi(d)$ distribution with $d$ degrees of freedom. Since the distribution $\chi(d)$ gets closer to a Dirac distribution when $d$ goes to infinity, one should use the highest dimension possible in order to obtain the degenerate version of the Fading Channel with Known Side Information where all the fading coefficients are equal to $1$, that is, the BIAWGNC. Unfortunately, the required division operator only exists in dimensions 1, 2, 4 and 8 (where it can be built from the algebraic structure of $\mathbb{R}$, $\mathbb{C}$, $\mathbb{H}$ and $\mathbb{O}$ respectively), so that it is not possible to use the above algorithm in arbitrary dimension.

\section{Reconciliation of Gaussian Variables: Implementation with LDPC Codes}
\label{sec:ldpcopt}
Low Density Parity Check (LDPC) codes (or Gallager codes) are linear error-correcting codes with a sparse parity check matrix. A good reference about general coding theory and LDPC codes is \cite{RIC08}. LDPC codes can be represented as bipartite graphs, one set of the nodes being the check nodes representing the set of parity-check equations which define the code; the other, the variable nodes which represent the elements of the codewords. Variables nodes and check nodes are connected through edges. LDPC codes are commonly used in telecommunications since they perform very close to Shannon limit and can be decoded with a fast iterative message-passing decoder called Belief Propagation (BP) (in such a decoding scheme, information is propagated between variable and check nodes that are connected by edges). These codes are designed for a given channel and a given SNR. The rate of a code is defined as the ratio between the information bits and the total number of transmitted bits on the channel. A low rate code is therefore a code with a lot of redundancy bits. Correcting errors at very low SNRs implies to design codes with low rates since adding redundancy allows to correct more errors.

A standard way to characterize LDPC codes is the probabilistic method: an ensemble of LDPC codes $\mathcal{C}$ is characterized by the node degrees and one proves that good codes occur with high probability within this ensemble. A specific code is simply drawn randomly from this set. Then one can modify the node degrees and their probabilities of occurence to improve the performance of the codes of the ensemble. A well known method to optimize LDPC codes for a given rate and a given channel is to use a genetic algorithm called \textit{Differential Evolution}. This method has been successfully applied for a wide range of channels: the Binary Erasure Channel (BEC) \cite{SHO00}, the BIAWGNC \cite{RIC01} and the Binary Symmetric Channel (BSC) \cite{ELK09}. The cost function that is maximized using this algorithm is defined as the threshold value for the channel (\textit{i.e.} the maximal value of the noise that can be corrected with a given code, e.g. the standard deviation $\sigma$ of the noise for the BIAWGNC or the probability of error $\epsilon$ for the BSC) and \textit{Discretized Density Evolution} is used to compute the threshold.

In CVQKD, we need low-rate and high-efficiency codes for the BIAWGNC since errors must be efficiently corrected at very low SNRs to increase the secure distance. \emph{Multi-edge-type LDPC codes} \cite{RIC04} give simple structures allowing to operate very close to Shannon limit at very low SNRs (for another construction of low rate LDPC codes refer to \cite{TIL10}). In the multi-edge setting, several edge classes are defined on the bipartite graph; then every node is defined by its number of sockets in each class. Whereas for standard LDPC ensembles the graph connectivity is constrained only by the node degrees, the multi-edge-type setting allows a greater control over the graph because only sockets of the same class can be connected together. Unlike standard LDPC ensembles, this framework provides for example the possibility to use degree-1 edges which improves significantly the threshold.

Every known reconciliation technique for CVQKD with a Gaussian modulation achieves an efficiency less than or equal to $90\%$ \cite{BLO05,LOD07,VAN06}. This efficiency parameter $\beta$ (defined by $\beta(s)=R / C(s)$ for a SNR $s$ where $R$ is the code rate used for the reconciliation and $C$ is the capacity of the Additive White Gaussian Noise Channel (AWGNC)) is critical since the asymptotic secure key rate in the reverse reconciliation scheme is given by $K=\beta I(x;y) - \chi(y;E)$, where both $I(x;y)$ (the mutual information between the two protagonists bit strings $x$ and $y$) and $\chi(y;E)$ (the Holevo information between the eavesdropper and the receiver's data) are large compared to $K$. One should especially pay attention to the dependency of $\beta$ on the SNR. In \cite{BLO05,LOD07,VAN06}, the good efficiency values are obtained only for SNRs higher than 1 which is incompatible with long distances. In \cite{LEV08}, a $90\%$ efficiency is obtained for a $0.5$ SNR which allows to extend the secure distance from 30 km to 50 km. In this paper, we obtain higher efficiencies for even lower SNRs which allows secure key distribution over longer distances.

Let us now review low rate LDPC codes with a good efficiency available in literature. In \cite{RIC04}, table IX, a $95.9\%$ efficiency, rate $1/10$ code for the BIAWGNC is described. This efficiency can be further improved through an optimization of the distribution coefficients as mentioned in \cite{RIC04}. Starting from the structure of this code we designed codes with lower rates and with higher asymptotic thresholds. Table \ref{tab:biawgnceff} sums up the performances of this original code together with our set of new multi-edge LDPC codes (the actual structure of the rate 0.02 code is described as an example in Appendix \ref{annex:ldpc}). In this table, $R$ is the rate of the considered code, $s_{DE}$ is the SNR threshold given by Discretized Density Evolution, $C_{th}$ is the theoretical channel capacity for this level of noise and $\beta_{DE}$ is the efficiency of the code. These results are valid in the asymptotic regime, \textit{i.e.} for codes of infinite length. However, the efficiency that is obtained with codewords of length $2^{20}$ is within $1\%$ of the asymptotic efficiency.
\begin{table}[t]
\centering
\begin{tabular}{|c|c|c|c|}
\hline
$R$ & $s_{DE}$ & $C_{th}$  & $\beta_{DE}$\\
\hline
0.1 & 0.156 & 0.10429 & 95.9\%\\
0.05 & 0.074 & 0.05144 & 97.2\%\\
0.02 & 0.029 & 0.02038 & 98.1\%\\
\hline
\end{tabular}
\caption{SNR asymptotic thresholds ($s_{DE}$) on the BIAWGNC, corresponding channel capacities ($C_{th}$) and efficiencies ($\beta_{DE}$) given by Density Evolution for low rate multi-edge LDPC codes of rate $R$.}
\label{tab:biawgnceff}
\end{table}
%\begin{center}
%\begin{tabular}{c|c|c|c|c}
%$R$ & $\sigma_{DE}$ & $s_{DE}$ & $C_{th}$  & $\beta_{DE}$\\
%0.1 & 2.535 & 0.156 & 0.10429 & 95.89\%\\
%0.05 & 3.67813 & 0.074 & 0.05144 & 97.20\%\\
%0.02 & 5.90723 & 0.029 & 0.02038 & 98.13\%\\
%\end{tabular}
%\end{center}
%\begin{center}
%\begin{tabular}{c|c|c|c}
%Rate & C_{th} & \sigma_{DE} & \beta_{DE}\\
%0.952 & 0.95313 & 0.4449 & 99.88\%\\
%0.5 & 0.50928 & 0.965 & 98.18\%\\
%0.43 & 0.44236 & 1.071 & 97.21\%\\
%0.1 & 0.10429 & 2.535 & 95.89\%\\
%0.05 & 0.05144 & 3.67813 & 97.20\%\\
%0.02 & 0.02038 & 5.90723 & 98.13\%\\
%\end{tabular}
%\end{center}

\subsection{Simulation Results with Rotations on $S^{1}$, $S^{3}$ and $S^{7}$}

Let us discuss the simulation results we obtained applying the multidimensional reconciliation scheme with the previous codes for a dimension $d=2$, $d=4$ and $d=8$, for the sign coding technique ($d=1$) and without using any additional information, \textit{i.e.} when we try to use a code designed for the BIAWGNC with a Gaussian modulation. 

Tables \ref{tab:rotationssnr} and \ref{tab:rotationseff} summarize the efficiencies we obtained with respect to the Gaussian channel capacity with our multi-edge LDPC codes for a block size of $2^{20}$. We obtained a quite high Frame Error Rate (FER) (about $1/3$) but a null Bit Error Rate (BER) on the blocks where the decoding succeeded. This means that concatenating our codes with very high rate codes like BCH codes to remove the residual errors (as was done in \cite{LOD07,FOS09}) is not necessary here.
%\begin{center}
%\begin{tabular}{c|c|c|c|c|c|c}
%$R$ & $\sigma$ & $\beta$ & $\sigma_{d=1}$ & $\beta_{d=1}$ & $\sigma_{d=2}$ & $\beta_{d=2}$\\
%0.1 & 1.92 & 57.85\% & 2.31 & 80.76\% & 2.43 & 88.65\%\\
%0.05 & 2.85 & 59.71\% & 3.50 & 88.34\% & 3.605 & 93.51\%\\
%0.02 & 4.6 & 60.04\% & 5.75 & 93.05\% & 5.85 & 96.27\%\\
%\end{tabular}
%\end{center}
%\begin{center}
%\begin{tabular}{c|c|c|c|c}
%$R$ & $\sigma_{d=4}$ & $\beta_{d=4}$ & $\sigma_{d=8}$ & $\beta_{d=8}$\\
%0.1 & 2.48 & 92.06\% & 2.495 & 93.10\%\\
%0.05 & 3.63 & 94.76\% & 3.65 & 95.77\%\\
%0.02 & 5.86 & 96.59\% & 5.87 & 96.92\%\\
%\end{tabular}
%\end{center}
\begin{table}[t]
\centering
\begin{tabular}{|c|c|c|c|c|c|}
\hline
$R$ & $s$ & $s_{d=1}$ & $s_{d=2}$ & $s_{d=4}$ & $s_{d=8}$\\
\hline
0.1 & 0.271 & 0.187 & 0.169 & 0.163 & 0.161\\
0.05 & 0.123 & 0.082 & 0.077 & 0.076 & 0.075\\
0.02 & 0.047 & 0.030 & 0.029 & 0.029 & 0.029\\
\hline
\end{tabular}
\caption{SNR thresholds on the BIAWGNC for low rate multi-edge LDPC codes (size $2^{20}$) using the multidimensional reconciliation scheme ($d=1,2,4,8$).}
\label{tab:rotationssnr}
\end{table}
\begin{table}[t]
\centering
\begin{tabular}{|c|c|c|c|c|c|}
\hline
$R$ & $\beta$ & $\beta_{d=1}$ & $\beta_{d=2}$ & $\beta_{d=4}$ & $\beta_{d=8}$\\
\hline
0.1 & 57.9\% & 80.8\% & 88.7\% & 92.1\% & 93.1\%\\
0.05 & 59.7\% & 88.3\% & 93.5\% & 94.8\% & 95.8\%\\
0.02 & 60.0\% & 93.1\% & 96.3\% & 96.6\% & 96.9\%\\
\hline
\end{tabular}
\caption{Efficiencies (w.r.t. the BIAWGNC capacity) for low rate multi-edge LDPC codes (size $2^{20}$) using the multidimensional reconciliation scheme ($d=1,2,4,8$)}
\label{tab:rotationseff}
\end{table}

Since the channel obtained with rotations is not exactly a BIAWGNC, the efficiencies $\beta$ are always lower than the efficiencies predicted by density evolution on the BIAWGNC. However, increasing the dimension $d$ of the rotations allows to get closer to the efficiency of the code on the BIAWGNC. This is expected since the norm of the input vector $u^d ||x^d||$ of the virtual channel follows a distribution $\chi(d)$ (where $d$ is the number of degrees of freedom), which gets closer to a Dirac when $d$ tends to infinity.

\begin{center}
\begin{figure}
\includegraphics[width=80mm]{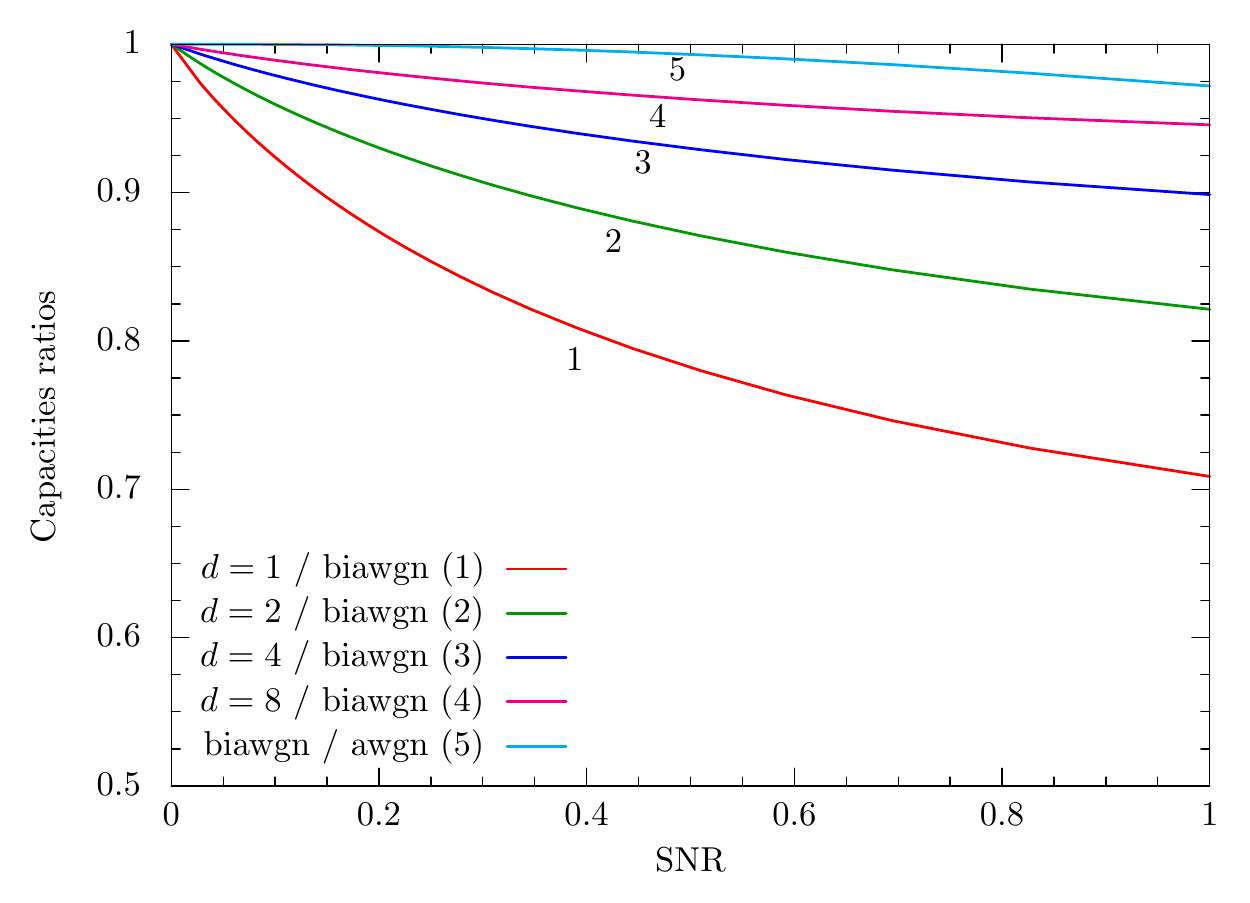}
\caption{(Color online) Ratios between the capacities of the multidimensional channels ($d=1,2,4,8$) and the BIAWGNC and between the BIAWGNC and the AWGNC with respect to the SNR}
\label{fig:capa}
\end{figure}
\end{center}

Figure \ref{fig:capa} compares the capacities of the BIAWGNC and the multidimensional virtual channels for $d=1,2,4,8$ as a function of the SNR.

%Without rotation:\\
%\begin{center}
%\begin{tabular}{c|c|c|c}
%Rate & \sigma & SNR & \beta\\
%0.5 & 0.53 & 3.560 & 56.40\%\\
%0.1 & 1.92 & 0.271 & 57.85\%\\
%0.05 & 2.85 & 0.123 & 59.71\%\\
%0.02 & 4.6 & 0.047 & 60.04\%\\
%\end{tabular}
%\end{center}
%With rotations on $S^{0}$:\\
%\begin{center}
%\begin{tabular}{c|c|c|c}
%Rate & \sigma & SNR & \beta\\
%0.5 & 0.615 & 2.644 & 61.80\%\\
%0.43 & 0.75 & 1.778 & 63.02\%\\
%0.1 & 2.31 & 0.187 & 80.76\%\\
%0.05 & 3.50 & 0.082 & 88.34\%\\
%0.02 & 5.75 & 0.030 & 93.05\%\\
%\end{tabular}
%\end{center}
%With rotations on $S^{1}$:\\
%\begin{center}
%\begin{tabular}{c|c|c|c}
%Rate & \sigma & SNR & \beta\\
%0.5 & 0.78 & 1.644 & 76.30\%\\
%0.43 & 0.90 & 1.235 & 77.30\%\\
%0.1 & 2.43 & 0.169 & 88.65\%\\
%0.05 & 3.605 & 0.077 & 93.51\%\\
%0.02 & 5.85 & 0.029 & 96.27\%\\
%\end{tabular}
%\end{center}
%With rotations on $S^{3}$:\\
%\begin{center}
%\begin{tabular}{c|c|c|c}
%Rate & \sigma & SNR & \beta\\
%0.5 & 0.865 & 1.336 & 85.69\%\\
%0.43 & 0.98 & 1.041 & 86.15\%\\
%0.1 & 2.48 & 0.163 & 92.06\%\\
%0.05 & 3.63 & 0.076 & 94.76\%\\
%0.02 & 5.86 & 0.029 & 96.59\%\\
%\end{tabular}
%\end{center}
%With rotations on $S^{7}$:\\
%\begin{center}
%\begin{tabular}{c|c|c|c}
%Rate & \sigma & SNR & \beta\\
%0.5 & 0.915 & 1.194 & 91.74\%\\
%0.43 & 1.025 & 0.952 & 91.49\%\\
%0.1 & 2.495 & 0.161 & 93.10\%\\
%0.05 & 3.65 & 0.075 & 95.77\%\\
%0.02 & 5.87 & 0.029 & 96.92\%\\
%\end{tabular}
%\end{center}

\subsection{Use of rotations in higher dimension spaces}
%With rotations on $ S^{15} $ :\\
%\begin{center}
%\begin{tabular}{c|c|c|c}
%Rate & \sigma & SNR & \beta\\
%0.5 & 0.935 & 1.144 & 94.27\%\\
%0.43 & 1.045 & 0.916 & 93.94\%\\
%0.1 & 2.51 & 0.159 & 94.14\%\\
%0.05 & 3.67 & 0.074 & 96.79\%\\
%0.02 & 5.90 & 0.029 & 97.89\%\\
%\end{tabular}
%\end{center}
%With rotations on $ S^{31} $ :\\
%\begin{center}
%\begin{tabular}{c|c|c|c}
%Rate & \sigma & SNR & \beta\\
%0.5 & 0.95 & 1.108 & 96.21\%\\
%0.43 & 1.055 & 0.948 & 95.19\%\\
%0.1 & 2.52 & 0.157 & 94.84\%\\
%\end{tabular}
%\end{center}
%High SNR, rate 0.952:
%\begin{center}
%\begin{tabular}{c|c|c|c}
%Dimension & sigma & SNR & \beta\\
%$S^0$ & - & - & -\\
%$S^1$ & 0.17 & 34.60 & 95.2\%\\
%$S^3$ & 0.29 & 11.89 & 95.30\%\\
%$S^7$ & 0.36 & 7.72 & 96.35\%\\
%$S^{15}$ & 0.40 & 6.25 & 97.57\%\\
%$S^{31}$ & 0.415 & 5.81 & 98.20\%\\
%$S^{63}$ & 0.425 & 5.54 & 98.70\%\\
%$S^{127}$ & 0.43 & 5.41 & 98.97\%\\
%$S^{255}$ & 0.435 & 5.28 & 99.26\%\\
%\end{tabular}
%\end{center}
As was explained in the previous section, the multidimensional reconciliation scheme is limited to dimensions $1, 2,4$ and $8$ because these are the only ones compatible with a division structure \cite{LEV08}. 

In \cite{LEV08}, the following construction applicable to arbitrary dimension $d$ is proposed. In the direct case, with the same notations as in paragraph \ref{sec:gaussianrec} (where Alice has a vector $\vx$, Bob a vector $\vy$, and Alice uses $(\vx,\myvector{r})$ to 'virtually' send $\vu$ to Bob), a random orthogonal transformation $Q$ on $\mathbb{R}^d$ is drawn according to the Haar mesure, then $Q$ is composed with the reflection $S$ across the mediator hyperplane of $\myvector{x'}=Q(\vx)$ and $\vu$. The resulting matrix $R=S \circ Q$ sends $\vx$ to $\vu$ and $\vy$ to a point close to $\vu$, because $R$ preserves the euclidean distance; $R$ is revealed by Alice and plays the same role as the vector $\myvector{r}$ in section \ref{sec:gaussianrec}. The randomization provided by $Q$ ensures that $R$ does not reveal more information on $(\vx,\vu)$ than the relation $R(\vx)=\vu$; in particular, all $\vu$ are equally likely given $R$.

$Q$ is built, for instance, as the orthogonal ('Q') part of the QR decomposition of a $d\times d$ matrix $G$ of Gaussian normalized random values. This method has complexity $\mathcal{O}(d^3)$. All other known methods to draw random orthogonal matrices have the same complexity.

\medskip
We propose a method that allows to reduce the complexity to $\mathcal{O}(d^2)$. Let us observe first that we have the choice of the encoding of $R$: we do not need to reveal it in matrix form. However, the encoding must not reveal anything about $\vu$ except that $R$ satisfies $R(\vx)=\vu$. For instance, with the first method, revealing separately $Q$ and $S$ instead of $R=S\circ Q$ is not a good idea since $S$ leaks information about $u$: indeed, in high dimension $d$, two random independent vectors are approximately orthogonal and therefore their mediator hyperplane forms and angle of about $\pi/4$ with either vector.

Let us examine first how an orthogonal transform $Q$ can be drawn according to the Haar measure with complexity $\mathcal{O}(d^2)$, using an adequate representation, the Householder decomposition. An orthogonal basis $\myvector{e}_1,\ldots,\myvector{e}_d$ is fixed. Let $E$ (resp. $F$) be the span of $\myvector{e}_1,\ldots,\myvector{e}_d$ (resp. $\myvector{e}_2,\ldots,\myvector{e}_d$).

If $d=1$, choose $+1$ or $-1$. If $d>1$, choose a random vector $\myvector{g}$ uniformly on $\mathcal{S}^{d-1}$, the unit sphere in $\mathbb{R}^d$ (it can be constructed as $\myvector{g} = \myvector{h}/||\myvector{h}||$ where $\myvector{h}$ has independent normalized Gaussian coordinates), and draw recursively a random orthogonal matrix $Q'$ of dimension $d-1$, viewed as a transform of $F$. $Q'$ is extended to $E$ by setting $Q'(\myvector{e}_1)=\myvector{e}_1$. Let $S$ be the reflection that sends $\myvector{e}_1$ on $\myvector{g}$, and define $Q = S \circ Q'$. $Q'$ is itself a composition of $d-1$ reflections in spaces of dimensions $d-1,\ldots,1$. Describing each reflection by its corresponding eigenvector for the eigenvalue -1, $Q$ is described by $d$ vectors of dimensions $d,d-1,\ldots,1$, for a total of $\frac{d(d+1)}{2}$ coefficients. The decomposition is unique. Note that $Q(\myvector{e}_1)=\myvector{g}$. 

This process can be adapted when a constraint $Q(\vx)=\vu$ is added, with $||\vx||=||\vu||$. If $d=1$, choose $+1$ or $-1$ depending on $\vx=\vu$ or $\vx=-\vu$. Assuming $d>1$, $\myvector{g}$ is chosen uniformly at random among unit vectors s.t.
\begin{equation}
\label{eq:constraint}
\vu \cdot \myvector{g}  = \vx \cdot \myvector{e}_1
\end{equation}
where $\cdot$ is the dot product. This relation is required for $Q$ to satisfy both $Q(\vx)=\vu$ and $Q(\myvector{e}_1) = \myvector{g}$. Starting from a Gaussian normalized vector $\myvector{h}$, $\alpha$ is chosen uniformly so that $(\myvector{h} + \alpha \vu) \cdot \vu = (\vx \cdot \myvector{e}_1) \times|| \myvector{h} + \alpha \vu||$ (this is a quadratic equation that has at least one solution except if $\myvector{h}, \vu$ span the same line, and  $\myvector{e}_1, \vx$ do not, which happens with probability 0). $\myvector{g} = \frac{\myvector{h} + \alpha \vu}{||\myvector{h} + \alpha \vu||}$ is computed in linear time and satisfies (\ref{eq:constraint}). 

For an arbitrary vector $\myvector{v}$, write its decomposition on $F$, $e_1$ as  $\myvector{v}=\myvector{v}_F+\myvector{v}_{F^{\perp}}$. $Q'$ is  drawn recursively, satisfying $Q'(\vx_F) = S(\vu)_F$. This is possible because $\vx \cdot \myvector{e}_1 = \vu \cdot \myvector{g} = \vu \cdot S(\myvector{e}_1) = S(\vu) \cdot \myvector{e}_1$ implies $\vx_{F^{\perp}} = S(\vu)_{F^{\perp}}$ and $||\vx_F|| = ||S(\vu)_F||$. Then as $Q'(e_1) = e_1$,  $Q'(\vx) = S(\vu)$. 

Define $Q=S\circ Q'$ as before: $Q(\vx)=S(Q'(\vx))=\vu$.

The algorithm still runs in $\mathcal{O}(d^2)$, and the decomposition does not reveal any side information because it is unique. Since the added constraint $(\ref{eq:constraint})$ is required for the relation $Q(\vx)=\vu$ to hold, one sees recursively that the process yields the correct distribution on $O_d$. Finally, given the $d$ reflection vectors, computing $Q(\vz)$ for any $\vz$ is also done in time $\mathcal{O}(d^2)$. Hence by revealing these vectors instead of $Q$ in matrix form, one gets the desired $\mathcal{O}(d^2)$ algorithm.

\medskip
Let us now consider the rate $1/2$ multi-edge LDPC code given in Table VI of reference \cite{RIC04}. The SNR threshold given by Discretized Density Evolution is $s^{*}=1.074$. The corresponding efficiency on the BIAWGNC is $98.2\%$.
When using a Gaussian modulation, table \ref{tab:highdim} shows the effect of the dimension $d$ on the efficiency $\beta$ of the reconciliation scheme. We can see that increasing the dimension above 8 when operating at a high SNR enables to  increase significantly the efficiency, and therefore the key rate in QKD applications.
\begin{table}[t]
\centering
\begin{tabular}{|c|c|c|}
%Dimension & sigma & $\beta$\\
%2 & 0.78 & 76.30\%\\
%4 & 0.865 & 85.69\%\\
%8 & 0.915 & 91.74\%\\
%16 & 0.935 & 94.27\%\\
%32 & 0.95 & 96.21\%\\
%64 & 0.955 & 96.86\%\\
\hline
$d$ & $s$ & $\beta$\\
\hline
2 & 1.644 & 76.3\%\\
4 & 1.336 & 85.7\%\\
8 & 1.194 & 91.7\%\\
16 & 1.144 & 94.3\%\\
32 & 1.108 & 96.2\%\\
64 & 1.097 & 96.9\%\\
\hline
\end{tabular}
\caption{SNR thresholds and channel efficiencies on the BIAWGNC for the rate $1/2$ multi-edge LDPC code in Table VI of ref. \cite{RIC04} with respect to the dimension of the multidimensional reconciliation scheme}
\label{tab:highdim}
\end{table}

\subsection{Dealing with a continuous range of SNR with puncturing, shortening and repetition}

We designed good efficiency codes for a finite set of rates so far; we are going to show how to deal with a continuous range of SNRs with this finite set.
Since we designed low rate codes with good efficiencies, we can apply the simple technique of repetition codes mentioned in \cite{LEV10}. It is shown that starting from a code of rate $R$ achieving an efficiency $\beta(s)$ for a SNR $s$ on the BIAWGNC, one can use a repetition scheme of length $k$ to build a new code of rate $R^{'}=R/k$ achieving an efficiency $\beta'$ for a SNR $s'=s/k$ given by
\begin{equation*}
\beta'(s/k)=\beta(s)\ \frac{\log_2(1+s)}{k\log_2(1+s/k)}
\end{equation*}
For example, using a repetition scheme of length $3$ with our code of rate $0.02$ and efficiency $98\%$ for a SNR of $0.03$, we can build a code of efficiency $\beta(0.01)=0.98\frac{\log_2(1.03)}{3\log_2(1.01)}=97\%$. We applied this technique with repetition factors of $2$ and $4$ with our code of rate $0.02$ to obtain the codes of rates $0.01$ and $0.005$ given in Table \ref{tab:setcodes}.

However, this technique allows a low efficiency loss only for very small SNRs. For higher SNRs, other techniques must be applied if we want to keep very good efficiencies. Puncturing and shortening for LDPC codes are a good way to adapt the rate of a code \cite{ELK10}. Let us start with a $(n,k)$ code, \textit{i.e.} a code of length $n$ with $n-k$ bits of redundancy; the rate is $R=k/n$. Puncturing consists in deleting a predefined set of $p$ symbols from each word, converting a $(n,k)$ code into a $(n-p,k)$ code. Shortening means deleting a set of $s$ symbols from the encoding process (or revealing $s$ message bits in addition to the syndrome in each codeword), converting a $(n,k)$ code into a $(n-s,k-s)$ code. With a combination of these techniques the rate obtained is
\begin{equation*}
R=\frac{k-s}{n-p-s}\quad .
\end{equation*}
The loss of efficiency incurred is small for small relative variations of the code rate. Typically, one can achieve a decrease of 5\% (though shortening) and an increase of 10\% (through puncturing) of the code rate with an efficiency loss smaller than 1\%.

\section{Practical use for a continuous-variable quantum key distribution system}
\label{sec:practical}
%\subsection{Optimal reconciliation parameters}
In this section, we apply the techniques developed in the previous sections to CVQKD in order to increase the secure distance achievable. We have to take into account that our quantum channel is Gaussian so that code efficiencies must be computed w.r.t. this channel capacity:
\begin{equation*}
\beta=\frac{R}{C_{AWGNC}}
\end{equation*}
where $R$ is the rate of the code and $C_{AWGNC}$ is the capacity of the AWGNC. As we can see on Figure \ref{fig:capa}, the capacity of the BIAWGNC is very close to the capacity of the AWGNC for small values of the SNR. We give the efficiencies we can achieve on the AWGNC for different SNRs in Table \ref{tab:setcodes}.
\begin{table}[t]
\centering
\begin{tabular}{|c|c|c|}
\hline
$R$ & $\beta$ & $s$\\
\hline
0.5 & 93.6\% & 1.097\\
0.1 & 93.1\% & 0.161\\
0.05 & 95.8\% & 0.075\\
0.02 & 96.9\% & 0.029\\
0.01 & 96.6\% & 0.0145\\
0.005 & 95.9\% & 0.00725\\
\hline
\end{tabular}
\caption{SNR thresholds and channel efficiencies on the AWGNC of the multi-edge LDPC codes mentionned in this paper.}
\label{tab:setcodes}
\end{table}

Our set of codes allows us to correct errors with an efficiency of about $95\%$ for some fixed low SNRs. Let us plot the secret key rate as a function of the SNR on Bob side for a given distance and assuming a fixed error correction efficiency $\beta$. This enables to determine for which particular SNR it is relevant to design error-correcting code in order to maximize the secret key rate. We do not consider in this paper finite-size effects \cite{LEV10b}, meaning that our figures represent the key rate in the regime of infinite block length. In order to take finite-size effects into account two approaches are possible: a theoretical one consists in improving the proofs and the bounds on the secret key rate \cite{BER11}, a more practical one  consists in designing systems with sufficient hardware stability in order to compute keys on large blocks. 

\begin{figure}
\includegraphics[width=90mm]{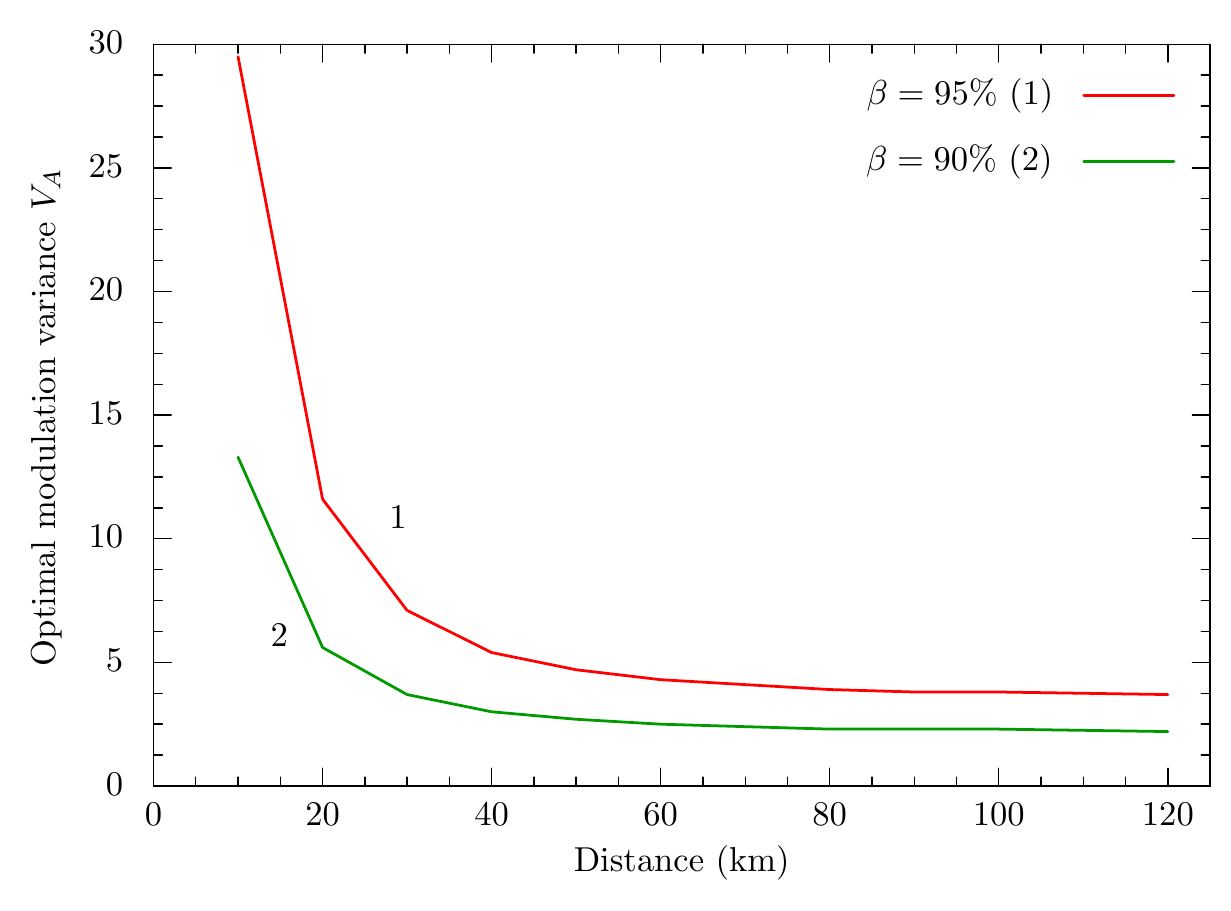}
\caption{(Color online) Optimal modulation variance with respect to the distance: $\eta=0.6$, $V_{elec}=0.01$, $\xi=0.01$, $\alpha=0.2dB/km$, $\beta=95\%$ and $\beta=90\%$ from top to bottom.}
\label{fig:distanceVAoptbeta9590}
\end{figure}

\begin{figure}
\includegraphics[width=90mm]{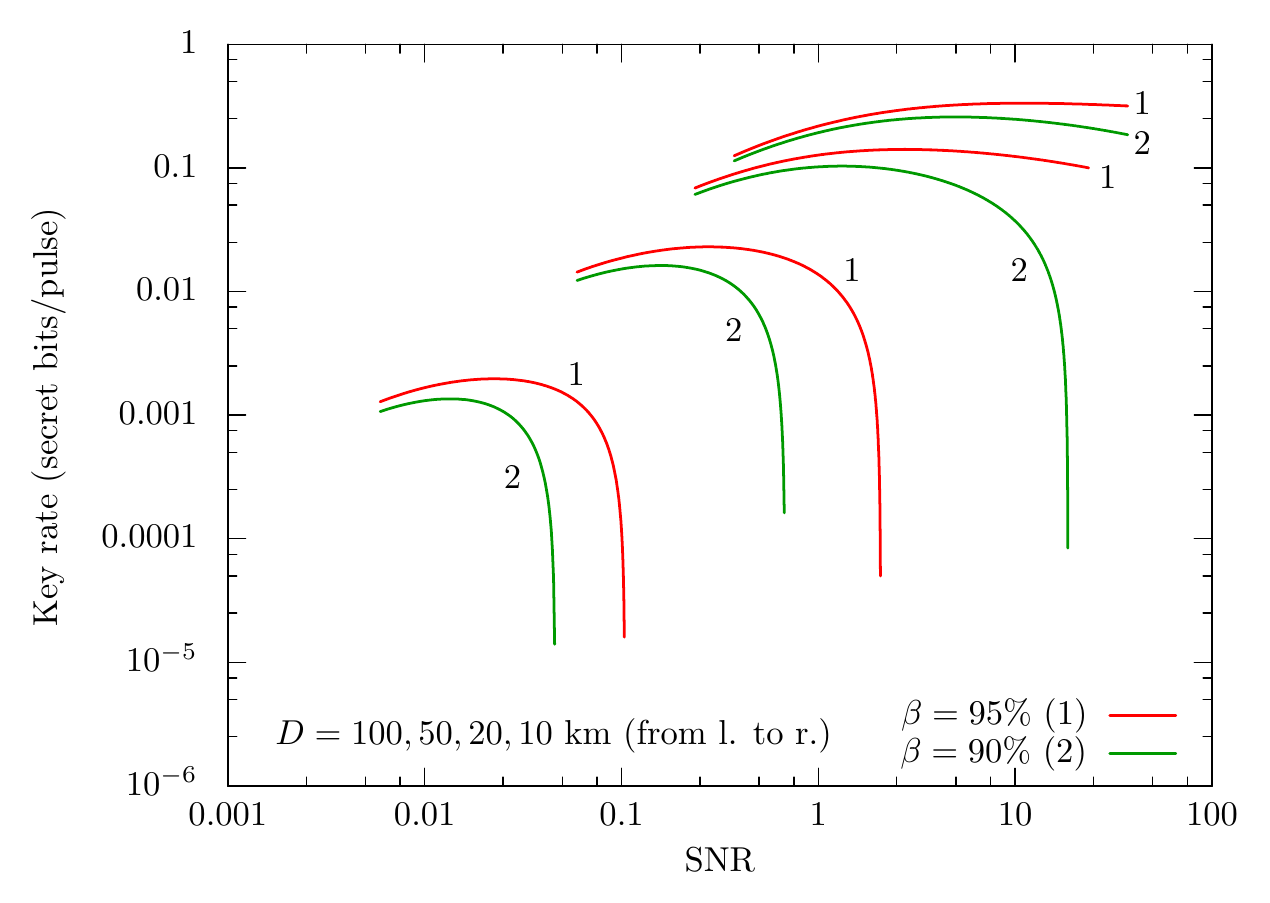}
\caption{(Color online) Secret key rate for collective attacks with respect to the SNR: $\eta=0.6$, $V_{elec}=0.01$, $\xi=0.01$, $\alpha=0.2dB/km$, $V_A \in \{1,100\}$, $\beta=95\%$ and $\beta=90\%$, $D=10,20,50,100$ km.}
\label{fig:keyratesnrbeta9590xi1}
\end{figure}

\begin{figure}
\includegraphics[width=90mm]{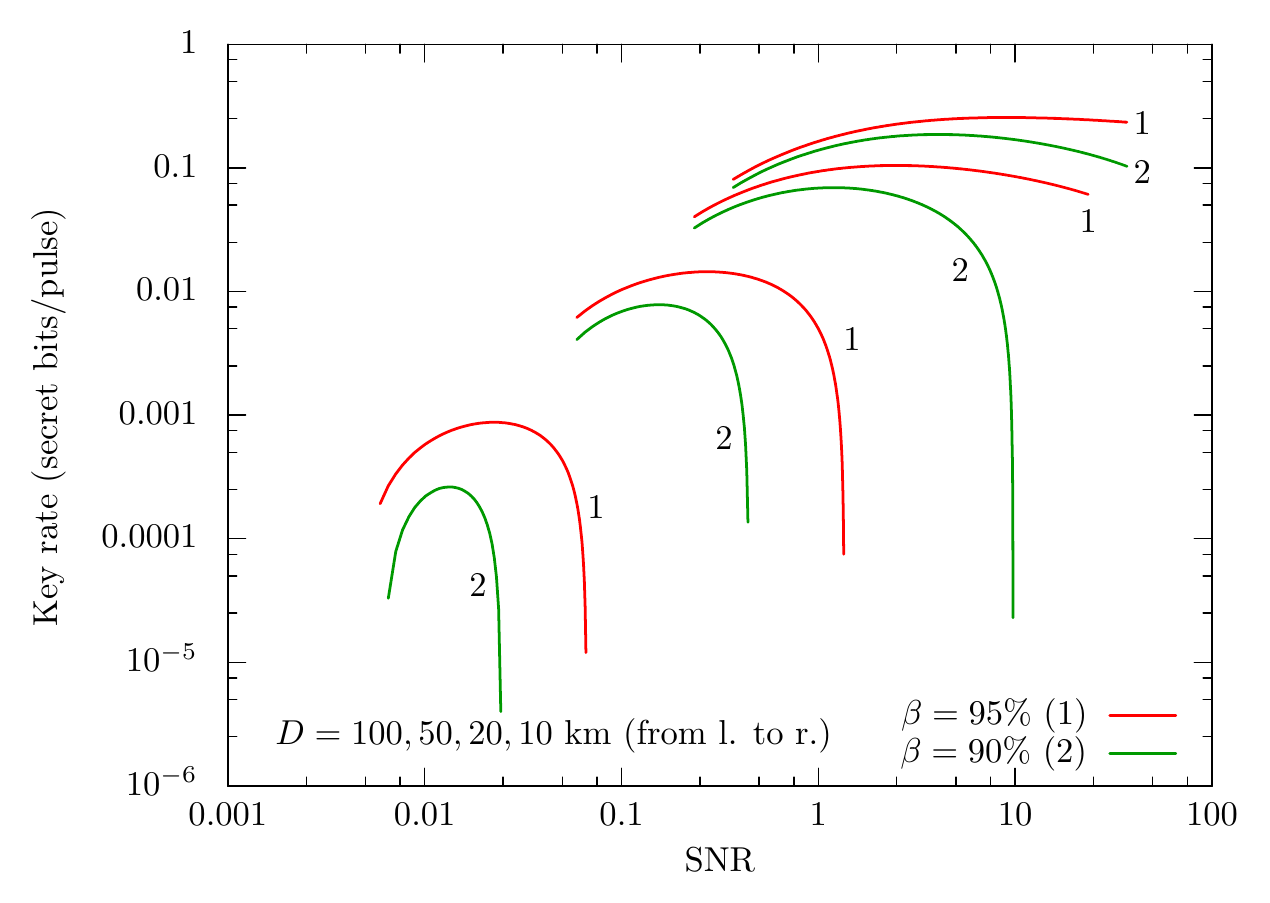}
\caption{(Color online) Secret key rate for collective attacks with respect to the SNR: $\eta=0.6$, $V_{elec}=0.01$, $\xi=0.04$, $\alpha=0.2dB/km$, $V_A \in \{1,100\}$, $\beta=95\%$ and $\beta=90\%$, $D=10,20,50,100$ km.}
\label{fig:keyratesnrbeta9590xi4}
\end{figure}

\begin{figure}
\includegraphics[width=90mm]{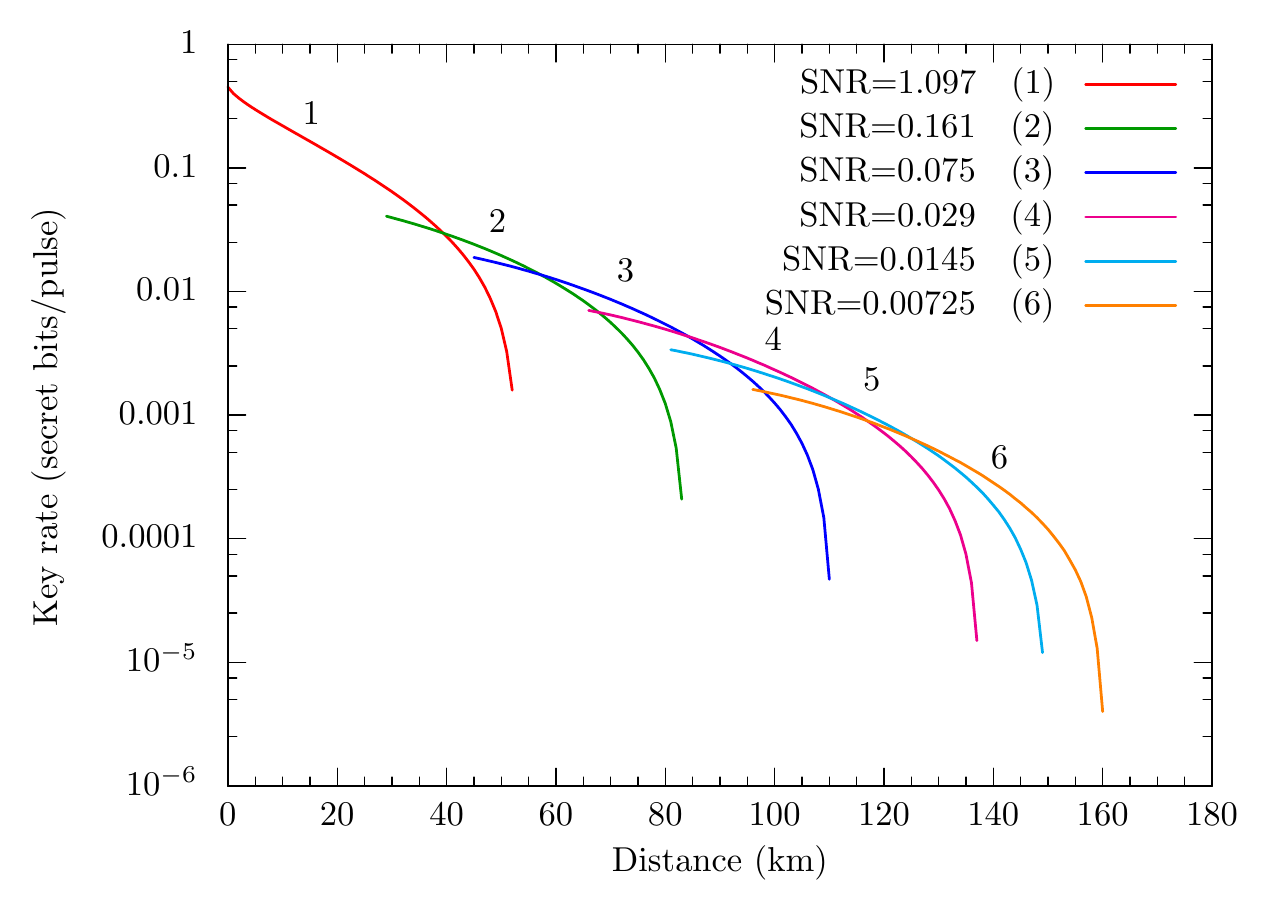}
\caption{(Color online) Secret key rate for collective attacks with respect to the distance: $\eta=0.6$, $V_{elec}=0.01$, $\xi=0.01$, $\alpha=0.2dB/km$, $V_A \in \{1,100\}$, $SNR=1.097,0.161,0.075,0.029,0.0145,0.00725$, $\beta=93.6\%,93.1\%,95.8\%,96.9\%,96.6\%,95.9\%$ from left to right.}
\label{fig:keyratedistance}
\end{figure}

The modulation variance is restricted within the interval $[1,100]$ (in shot noise units) since lower values make the experimental setup much more complex. Indeed, a very low modulation variance is not compatible with brighter synchronization and phase tracking signals, because of the limited extinction ratios of the optical modulators (30dB for the most common models). An attenuation of 0.2dB/km is assumed. The homodyne detection  efficiency is set to $0.6$, and a value of $1\%$ of the shot noise is taken for the electronic noise of the homodyne detection \cite{LOD07,FOS09}. A conservative value of $4\%$ of the shot noise as in the European project SECOQC (Secure Communication based on Quantum Cryptography) \cite{FOS09} is used for the excess noise in Figure \ref{fig:keyratesnrbeta9590xi4} while a more optimistic figure of $1\%$ is used in Figures \ref{fig:distanceVAoptbeta9590}, \ref{fig:keyratesnrbeta9590xi1} and \ref{fig:keyratedistance}. This second value is also typical of a realistic CVQKD system \cite{LOD07}.


Figure \ref{fig:distanceVAoptbeta9590} shows the optimal variance modulation on Alice side with respect to the key rate as a function of the distance. Achieving a good reconciliation efficiency at any SNR allows to work with a high modulation variance. This compares favorably to previous schemes with a discrete modulation which require modulation variances $10$ times lower than the ones shown here.

Figure \ref{fig:keyratesnrbeta9590xi1} and \ref{fig:keyratesnrbeta9590xi4}, plotted respectively for an excess noise of $1\%$ and $4\%$ of the shot noise, show that an improvement on the reconciliation efficiency yields at any distance a wider range of SNR with a close-to-optimum secret key rate. Conversely, the range of distances where a given error-correcting code working close to its threshold SNR can be used to get an almost optimal key rate is increased. 

Given these large distance ranges where an error-correcting code is usable, it becomes feasible to use a small family of error-correcting codes to perform the reconciliation step at \emph{any} distance and \emph{without} using rate-tuning techniques such as puncturing or shortening. Figure \ref{fig:keyratedistance} shows the key rate and the maximum secure distance obtained with this simple approach and the codes of Table \ref{tab:setcodes}. 
With an excess noise of $1\%$ of the shot noise, a secure distance above 150 km is obtained (with an excess of noise of $4\%$ and the same codes, the secure distance is above 140 km). This is a significant improvement over previous reconciliation techniques since a reconciliation efficiency of $90\%$ for a SNR of $0.5$ only allows a secure distance of about 50 km with a Gaussian modulation \cite{LEV08}.

\section{Conclusion}
We designed high-efficiency error-correcting codes allowing to distribute secret keys with a continuous-variable quantum key distribution system using a Gaussian modulation over long distances. Our results give a secure distance above 150 km against collective attacks (in the asymptotic regime) and can be implemented with only software modifications in the experimental setups of \cite{LOD07} and \cite{FOS09}.

\begin{acknowledgments}
This research was supported by the ANR, through the FREQUENCY and HIPERCOM projects, and by the European Union, through the FP7 project Q-CERT and ERC Starting Grant PERCENT.
\end{acknowledgments}

\bigskip 
\appendix
\section{A rate 1/50 multi-edge LDPC code ($\sigma^{*}=5.91$ on the BIAWGNC)}

\label{annex:ldpc}
Below is the description of a multi-edge LDPC ensemble of codes of rate $R=0.02$. The left half of the array describes the multidegree distributions of variable nodes, and the right half the distribution of check node multidegrees. $m$ stands for a multidegree distribution of probability $\nu_m$ at the variable nodes and $\mu_m$ at the check nodes. For instance, with probability 0.0225, a variable node has multidegree $[2,\ 57,\ 0]$, \textit{i.e.} it has 2 sockets for edges of type 0, 57 sockets for edges of type 1, and no socket of type 2. Check node probabilities sum to $1-R=0.98$ since there is 0.98 check node for 1 variable node.

$$
\begin{array}{|c|c||c|c|}
\hline
 \nu_{m} & m &  \mu_{m} & m\\ 
\hline
 0.0225  & 2 \quad 57 \quad 0 & 0.010625 & 3 \quad 0 \quad 0\\
 0.0175 & 3 \quad 57 \quad 0 & 0.009375 & 7 \quad 0 \quad 0\\
 0.96  & 0 \quad 0 \quad 1 & 0.6 & 0 \quad 2 \quad 1\\
 &  & 0.36 & 0 \quad 3 \quad 1\\
\hline
\end{array}
$$

\bibliography{biblio}
	
\end{document}